\documentstyle[twocolumn,prl,aps]{revtex}
\tighten
\begin{document}
\draft
\title{Resonance lineshapes in quasi-one-dimensional scattering}
\author{Jens U.~N{\"o}ckel and A.~Douglas Stone}
\address{
Applied Physics, Yale University, P.O. Box 208284,
Yale Station, New Haven CT 06520-8284\\{(\rm Submitted 24 November 1993)}\\
\parbox{14 cm}{\smallskip\rm
An S matrix approach is developed to describe elastic scattering resonances of
systems where the scattered particle is asymptotically confined and
the scattering potential lacks continuous symmetry.
Examples are conductance resonances in microstructures or transmission
resonances in waveguide junctions.  The generic resonance is shown
to have the asymmetric Fano lineshape.  The asymmetry parameter $q$ is
independent of coupling to the quasi-bound level implying a
scaling property of the resonances which can be tested in transport
experiments.}}
\maketitle

Although the symmetric Breit-Wigner peak \cite{breit}
is the most common
resonance lineshape observed in atomic and nuclear scattering, it has been
known for some time that the most general resonant lineshape is
described by the asymmetric Fano function:
\begin{equation}\label{fanofunc}
f(\epsilon)=\frac{(\epsilon+q)^2}{\epsilon^2+1}.
\end{equation}
Here, $\epsilon=2(E-E_0)/\Gamma$ is the (dimensionless) energy from
resonance,
$\Gamma$ is the resonance width, and $q$ is the asymmetry parameter.
Strongly asymmetric Fano lineshapes are familiar for
inelastic autoionizing resonances in atoms\cite{fano}; however Simpson
and Fano\cite{simpson} explicitly noted their occurence in
spherically-symmetric {\it elastic} scattering in 1963. In fact
this lineshape is implicit in much earlier work in nuclear scattering
where strong asymmetries were measured in elastic neutron scattering \cite
{adair,blatt}.  To be precise one finds
\cite{simpson} that the elastic scattering partial cross-section near
a resonance in angular momentum channel $l$ (neglecting spin) has the form
\begin{equation}\label{partcross3d}
\sigma_l=\frac{\lambda^2}{2\pi}\,(2l+1)\,\sin^2{\bar\theta}_l\,
\frac{(\epsilon-\cot{\bar\theta}_l)^2}{\epsilon^2+1},
\end{equation}
with $\lambda$ denoting the particle wavelength and
${\bar\theta}_l$ the background phase shift in the
absence of the resonance. Hence $\sigma_l$ is proportional to the Fano function
with $q=-\cot{\bar\theta}_l$. The only assumption needed is that
${\bar\theta_l}$ is roughly constant with energy across the resonance.
Note that in the limit ${\bar\theta_l} \to 0$ ($q \to \infty$)
of small background phase shift one recovers the Breit-Wigner lineshape;
whereas $q \to 0$ yields a symmetric anti-resonance (BW dip).
The characteristic asymmetric Fano lineshape occurs for intermediate
values of $q$ and in the inelastic case has the well-known interpretation
of interference between transition amplitudes from a bound initial state
to an unbound final state either directly or via a quasibound intermediate
state.  A similar interpretation applies
for asymmetric elastic resonances in which scattering via a
quasi-bound level interferes with direct (potential) scattering.
An important feature of elastic Fano resonances is that the asymmetry
parameter depends only on the background phase-shift and not on the strength
of the coupling to the quasi-bound level; this is not true for inelastic
autoionizing resonances.  Note that in the spherically symmetric case
$\sigma_l$ varies between zero and $\lambda^2 (2l+1)/2\pi$, the minimum and
maximum values allowed by unitarity of the S matrix.

In several recent theoretical studies of transport properties of
microstructures
\cite{buttiker84,ko,schult,baranger,porod,bagwell,noeckel2,tekman},
asymmetric elastic scattering conductance resonances have been encountered.
Typically an impurity or geometric feature
of the microstructure leads to the formation of a quasi-bound state
which is degenerate with the continuum of propagating
states.  Although resonance lineshapes were calculated in several
specific models, and the analogy to Fano resonances was discussed in an
interesting paper by Tekman and Bagwell\cite{tekman}, no general proof
that these resonances have the Fano lineshape
(analogous to that for spherically-symmetric continuum scattering)
has been given.  Unlike atomic or nuclear scattering in 3D,
for electrons in microstructures or photons in waveguides
one must consider
geometries in which the scattering occurs between different
propagating channels in leads which confine the particle asymptotically.
Thus, we may assume at most discrete rotational symmetry, and the lineshape
in the absence of spatial symmetry and/or time-reversal symmetry
will be of great interest.
We use the term quasi-one-dimensional (Q1D) scattering to emphasize the
one-dimensional nature of the asymptotic motion in such systems.
We note that a strong magnetic field also confines
electrons asymptotically even in 3D, so our results are relevant
to 3D heterostructures as well.
We obtain the following results:
1) The Fano function describes the
generic lineshape for Q1D resonant scattering in the presence and absence
of discrete symmetries.
2) The asymmetry parameter $q$ again is
independent of the coupling strength. This implies a scaling property
of the resonances which we test on an example below.  3) Symmetric
two-probe structures in the single-subband regime
always show the maximum variation in the transmission through resonance
allowed by unitarity.  In the absence of
symmetries unit reflection is maintained but unit transmission is lost.
Finally, Fano resonances in
microstructures have not yet been clearly seen experimentally
(although some data taken
on quantum point contacts is suggestive \cite{McEuen});
we propose below a potentially more tractable experiment for observing
such resonances in 3D heterostructures.

Our discussion of Q1D resonance lineshapes
exploits the fundamental relationship between resonances and poles
of the S matrix in the complex energy plane, which is also the basis of
Eq.\ (\ref{partcross3d}). Consider a Q1D structure with $N$ leads and
various propagating subbands per lead, the total number of subbands in
all the leads being $M$.
An arbitrary scattering state is then specified
by $M$-vectors ${\bf I}$ and ${\bf O}$ for the amplitudes of the
incoming and outgoing waves in each subband of each lead.
The S matrix is defined by ${\bf O}=S\,{\bf I}$,
and current conservation requires it to be unitary for real energies.
The wavefunction of a metastable state at energy $E_0$ with
decay width $\Gamma$ corresponds to a nonzero ${\bf O}$ with
vanishing ${\bf I}$ at a complex energy \cite{blatt,taylor}
${\bar E}\equiv E_0-i\Gamma/2$.
For this to be possible, $S^{-1}$ must be singular at this energy,
which implies that
one of the eigenvalues of $S$ has a pole
at ${\bar E}$. As is usually done, we assume the pole to be simple, and
only a single eigenvalue to be resonant (nondegenerate resonance).
With the constraint that this eigenvalue $\lambda_j(E)$ be
unimodular for real energies, one can approximate
\cite{taylor}
\begin{equation}\label{eigenapp}
\lambda_j(E)\approx e^{2i{\bar\theta}_j}
\frac{E-{\bar E}^*}{E-{\bar E}},
\end{equation}
where ${\bar\theta}_j$ is a real constant. This is accurate provided
that the distance to other resonance poles is much larger than
$\Gamma$. Introducing the eigenphase,
$\lambda_j=e^{2i\theta_j}$, one arrives at
\begin{equation}\label{phase}
\theta_j\approx{\bar \theta}_j-\arctan\frac{\Gamma/2}{E-E_0}.
\end{equation}
Thus, $2\theta_{j}$ varies by $2\pi$ on resonance.
To deduce the S matrix elements from a knowledge
of its eigenvalues, one has to specify the unitary transformation that
diagonalizes $S$. For the case of spherically symmetric 3D (and 2D)
scattering this
transformation is fixed because there is a complete set of quantum
numbers ($E$, $l$ ,$m$) that are conserved by $S$, leading to the
expression in Eq.\ (\ref{partcross3d}).
In the absence of unitary and antiunitary (time-reversal)
symmetries an approach which avoids
specifying the diagonalizing transformation (as e.g. in multi-channel
scattering theory \cite{landau,taylor}) is to make the
ansatz that each S matrix element $S_{mn}$ will itself
exhibit a resonance denominator as in Eq.\ (\ref{eigenapp}).
\begin{equation}\label{multichan}
S_{mn}=S_{mn}^{b}-i\frac{\gamma_m\,\delta_n^*}{E-E_0+i\Gamma/2},
\end{equation}
where $S^{b}$ is the background scattering matrix in the absence of the
resonance and ${\vec\gamma}$, ${\vec\delta}$ are complex vectors which must
satisfy ${\vec\gamma}=S^{b}{\vec\delta}$ and
$\vert\vec\gamma\vert^2=\vert\vec\delta\vert^2=\Gamma$ so that S is unitary.
$\vert\gamma_m\vert^2$, $\vert\delta_n\vert^2$ are then interpreted
as partial widths for leaving and entering
the resonance \cite{taylor}.

A simplified version of Eq.\ (\ref{multichan}) which assumes the background
S-matrix to be diagonal is often used\cite{landau} and in particular was
employed by B\"uttiker \cite{buttiker88}
to derive the lineshape of resonances in Q1D structures. However this further
approximation always leads to symmetric resonance lineshapes.
On the other hand a general
unitary $N\times N$ matrix is specified by $N^2$ real parameters,
and clearly
Eq.\ (\ref{multichan}) contains more than $N^2$ parameters since
$S^{b}$ is itself a general unitary matrix. It turns out that the
parameterization
in Eq.\ (\ref{multichan}) underconstrains the S-matrix, allowing
lineshapes that cannot arise in reality
from a nondegenerate, simple and isolated resonance pole.
For instance, according to Eq.\ (\ref{multichan}) a
$2\times 2$ S-matrix with $S^b_{12}=S^b_{21}=1$ gives rise to a non-vanishing
transmission $S_{12}$ if $\gamma_2\ne 0$.  However we show
below that transmission zeros are present in the most general
two-probe ($2 \times 2$) Q1D resonant S-matrix.
(Exact transmission zeros have been found in various model calculations
cited above and according to an argument in Ref.\ \cite{gurvitz} are
expected to be very robust in Q1D systems).

To begin we consider a Q1D geometry with $N$ leads in the single-subband regime
which is invariant under the {\em finite} rotation
group $C_N$ (an example being the symmetric cross junction \cite{schult}).
The symmetry will then determine the transformation which diagonalizes the
S-matrix in close analogy to the 3D case.
To find the eigenbasis we note that
the $N$ one-dimensional irreducible representations of $C_N$ have the character
system
\begin{equation}\label{character}
\chi^{(q)}(p)=e^{-2\pi ipq/N},\qquad(p,\,q=1,\ldots N).
\end{equation}
Here, $p$ labels the elements $R_p$ of the rotation group, and $q$
enumerates the representations.
{}From the $N$ degenerate scattering states corresponding to an
incoming electron
in exactly one of the leads, we can form symmetrized eigenfunctions by
taking the incoming waves as ${\bf I}^{(q)}$ with components
\begin{equation}\label{components}
I^{(q)}_p=\frac{1}{\sqrt{N}}\chi^{(q)}(p)^*.
\end{equation}
For any rotation $R_p$, one then has
$R_p\,{\bf I}^{(q)}=\chi^{(q)}(p)\,{\bf I}^{(q)}$. But since $R_p$
leaves the system invariant, a rotation of the incoming wave amplitudes
$R_p\,{\bf I}$ leads to outgoing waves $R_p\,{\bf O}$. For the
symmetrized waves this means
\begin{equation}\label{rot}
R_p\,{\bf O}=S\,R_p\,{\bf I}^{(q)}=\chi^{(q)}(p)\,S\,{\bf I}^{(q)}=
\chi^{(q)}(p)\,{\bf O}.
\end{equation}
Consequently, ${\bf O}$ transforms under the rotations in the same
way as ${\bf I}^{(q)}$. Since the representations are one-dimensional,
it follows that ${\bf O}\propto {\bf I}^{(q)}$,
so that the ${\bf I}^{(q)}$ are an eigenbasis of $S$.
The unitary transformation relating the matrix elements of $S$ between
incoming
and outgoing waves in leads $m$ and $n$ to the diagonal elements
$\lambda_j$ is then given by
\begin{eqnarray}
S_{mn}&=&\frac{1}{N}\sum\limits_{j=1}^{N}
\chi^{(m)}(j)^*\,\lambda_j\,\chi^{(n)}(j),\\
&=&\frac{1}{N}\sum\limits_{j=1}^{N}e^{2i\theta_j}\,e^{2\pi i(m-n)j/N}.
\label{usmatrix}
\end{eqnarray}
This is an exact expression for the multiprobe scattering amplitude.
Now assume that $\theta_1$ is the resonant eigenphase
while all other $\theta_j$ are slowly
varying with energy. Then we abbreviate the sum over the nonresonant
eigenvalues by
\begin{equation}
\sum\limits_{j=2}^{N}e^{2i\theta_j}\,e^{2\pi i(m-n)j/N}\equiv
\rho_{mn}\,e^{2i{\tilde\theta}_{mn}}
\label{usmatrix1}
\end{equation}
where $\rho\le N-1$. With this one obtains for the scattering
probabilities
\begin{equation}
\left|S_{mn}\right|^2=\left(\frac{\rho_{mn}+1}{N}\right)^2
-4\frac{\rho_{mn}}{N^2}\sin^2({\tilde\theta}_{mn}
-\theta_1),\label{transprob}
\end{equation}
which implies that each matrix element of $S$ varies in modulus between
$(\rho_{mn}-1)/N$ and $(\rho_{mn}+1)/N$ when crossing the resonance.
{}From this we see immediately that the $S_{mn}$ {\em must} take on any
value between and including $0$ and $1$ if $N=2$ (necessarily $\rho_{mn}=1$),
but {\em cannot} reach all values allowed by unitarity when $N>2$. To
find the lineshape resulting from Eq.\ (\ref{transprob}), we insert
Eq.\ (\ref{phase}) for $\theta_1$ and get
\begin{eqnarray}
\left|S_{mn}\right|^2&=&\left(\frac{\rho_{mn}+1}{N}\right)^2\nonumber\\
&&-4\frac{\rho_{mn}}{N^2}
\sin^2\left[{\tilde\theta_{mn}}-{\bar\theta_1}
+\arctan\frac{\Gamma/2}{E-E_0}\right]\nonumber\\
&\equiv&\left(\frac{\rho_{mn}+1}{N}\right)^2\nonumber\\
&&-4\frac{\rho_{mn}}{N^2}
\vert t^b_{mn}\vert^2\,\frac{\left(E-E_0+\Delta_{mn}\right)^2}
{(E-E_0)^2+\Gamma^2/4}.\label{transpole1}
\end{eqnarray}
Here, we have identified
$\vert t^b_{mn}\vert^2=\sin^2({\tilde\theta}_{mn}-{\bar\theta}_1)$
as the (slowly
varying) transmission in the absence of the resonance, and introduced
the energy shift
$\Delta_{mn}=\frac{1}{2}\Gamma\cot({\tilde\theta}_{mn}-{\bar\theta_1})$.
This is the Fano
lineshape, Eq.\ (\ref{fanofunc}), with $\epsilon=(E-E_0)/(\frac{1}{2}
\Gamma)$
and $q=\Delta_{mn}/(\frac{1}{2}\Gamma)=
\cot({\tilde\theta}_{mn}-{\bar\theta_1})$,
superimposed on a constant background.
Owing to Eq.\ (\ref{usmatrix1}), the reflection $\vert S_{nn}\vert^2$
can reach unity only if all nonresonant eigenvalues are the same,
in which case $\rho_{nn}=N-1$. But this is
automatically satisfied in the two-probe structure, $N=2$, where there
is only one nonresonant eigenphase, ${\tilde\theta_{nn}}=\theta_2$.
In that special case, the transmission becomes
\begin{equation}
T=1-\left|S_{11}\right|^2=
\vert t^b\vert^2\,\frac{\left(E-E_0+\Delta\right)^2}
{(E-E_0)^2+\Gamma^2/4},\label{transpole}
\end{equation}
which goes through zero and one.

We now consider the lineshape in the absence of symmetry for the
multi-probe, multi-subband case.
Let $U$ be the unitary matrix that diagonalizes $S$,
and $\theta_{1}$ be the resonant
eigenphase, all other $\theta_k$ being only weakly energy dependent.
The reflection amplitude is
\begin{equation}\label{reflamp}
S_{nn}=e^{2i\theta_1}\,\vert U_{1n}\vert^2
+\sum\limits_{k>1}e^{2i\theta_k}\,\vert U_{kn}\vert^2.
\end{equation}
where the index $k$ runs over leads and sub-bands.

Eq.\ (\ref{reflamp}) is completely general and implies that
$\vert S_{nn}\vert^2$
reaches unity if and only if all the nonresonant eigenvalues
are identical. As a consequence, the resonant transmission
$T$ of a two-probe
structure in the single subband regime {\em still} has a zero at
some energy even in the absence of symmetries.
This conclusion requires no special assumptions about
$U$; however some further information is needed to obtain the lineshape.
In the symmetric case
Eq.\ (\ref{components}) implies that the eigenvectors have a constant
modulus of $1/\sqrt{N}$. If we assume that these moduli (corresponding
to the incident fluxes in each lead) are only weakly
dependent on energy even in the absence of symmetries, the two-probe
transmission becomes
\begin{equation}\label{transasym}
T=(1-a^2)\vert t^b\vert^2\,\frac{\left(E-E_0+\Delta\right)^2}
{(E-E_0)^2+\Gamma^2/4}
\end{equation}
where the constant $a^2<1$. Thus we recover the Fano lineshape
of Eq.\ (\ref{transpole}) with a reduced prefactor; this implies that
perfect reflection still occurs at resonance, but not perfect transmission.

Although the Fano function, Eq.\ (\ref{fanofunc}), is the
generic lineshape for Q1D scattering it is never seen in
purely 1D resonant tunneling because in this case the background transmission
and lifetime of the metastable state are not independent.
Well-defined resonances with small $\Gamma$
require low background transmission, which simply gives the
Breit-Wigner lineshape. On the other hand, if $\Gamma$ is not small then
the assumption of constant ${\tilde\theta}$ needed
in Eq.\ (\ref{transpole}) is not valid over the width of the
resonance.  In a Q1D system, on the other hand, an electron entering the region
where the quasi-bound state is localized does not necessarily enter that
state itself, because the existence of a second scattering
channel allows resonant and nonresonant transmission to occur in
parallel as two distinct processes. The background transmission can still
be large even if the coupling to the quasi-bound level (which determines
$\Gamma$) is small, (e.g. due to approximate symmetry). Since the energy shifts
$\Delta$ in Eqs.
(\ref{transpole}),(\ref{transasym}) are proportional to $\Gamma$,
the asymmetry parameter $q$ defining the lineshape is actually
{\it independent} of $\Gamma$. If $\Gamma$ can be
varied while $\vert t^b\vert^2$ is roughly constant across resonance,
a series of Fano lineshapes will be obtained
which can be collapsed onto a curve characterized by a single
asymmetry parameter $q$ by rescaling the energy axis.  This scaling
property may be tested for the first time in transport experiments.

We have explored several different systems which might exhibit Fano
resonances when appropriately perturbed to create a quasi-bound
level in the continuum\cite{noeckel2,noeckel4}.
These include quantum point contacts, 2D electron gas
systems with an in-plane magnetic field, and magnetotransport in 3D
heterostructures. Here we focus on the 3D case because epitaxial
heterostructures may be grown with atomic precision so the effects of
disorder are minimized.  We consider
a quantum well of finite depth sandwiched between bulk emitter and
collector regions with a magnetic field ${\bf B}$ oriented
normal to the layers (see inset to Fig. \ref{fig1}).
As noted, the motion along ${\bf B}$
reduces to a Q1D problem due to Landau quantization in the
transverse plane (the Landau level index plays the role of the sub-band
index).  The well gives rise to a finite sequence of
bound states which repeats itself below each Landau level threshhold
$E_n$. The number and binding energies of these bound states
depend on well parameters which may be controlled, hence for $n>1$
they may lie degenerate with the continuum of propagating states associated
with lower LL's.  For precisely normal magnetic fields the LL wavefunctions
of bound and continuum states
are orthogonal and no resonance occurs, but a small tilt angle
will induce inter-LL coupling at the interface and
Fano resonances whose width may be tuned by varying the tilt angle
$\alpha$ (see Fig. \ref{fig1}).

The experiment must be done at low temperature (so that thermal broadening
does not distort the intrinsic lineshape) and in the linear response
regime.  Thus we propose tuning the Fermi energy, $E_F$
through resonance by varying the magnitude of the magnetic field, B.
Above some threshhold field $B_0$, $E_F$ will enter
the lowest LL and lie slightly below $E_2$.
As $B$ is increased further, $E_F$ approaches the bottom of the lowest LL,
$E_1$. The quasibound levels also
shift as a function of $B$ but $E_F$ will
cross all resonances between $E_2$ and $E_1$ at some $B$.
The resulting
transmission curves as a function of $B$ (obtained from simulations)
are shown in Fig.\ \ref{fig1}
for a particular resonance at various small tilt angles. The scaling
property of the resonances is seen to be
very well satisfied in this system (Fig. 1b).
Discussions and simulations of this and other possible experimental geometries
will be presented in detail elsewhere\cite{noeckel4}.

We acknowledge R.~Wheeler for the important suggestion of tuning
through resonance with a magnetic field.  We also thank
M.~B\"uttiker, M.~Reed and R.~Adair for helpful discussions.
This work was supported by ARO grant no.\ DAAH04-93-G0009.

\narrowtext
\begin{figure}
\caption{\label{fig1}
(a) Transmission of a finite well as a function of magnetic field $B$ in units
of $B_0$. The resonances correspond, from left to right, to tilt angles of
$\protect{\sin\alpha}=0.02$, $0.01$, $0.005$ (curves offset for clarity).
With $\omega\equiv eB_0/mc$, the well has depth
$V=2\,\hbar\omega$ and width $L=3.5\protect\sqrt{\hbar/m\omega}$.
(b) The same resonances plotted
in reduced units, $\epsilon$ being defined as below Eq.\ (
\protect\ref{transpole}); the Fano asymmetry parameter is $q=0.887$. Over the
width of these narrow resonances, $E_F$ is proportional to $B$ while $\vert
t^b\vert^2$ and $\Gamma$ are independent of $B$.
Inset: effective potential in the growth direction.}
\end{figure}

\end{document}